\documentclass[showpacs,10pt,twocolumn,prb]{revtex4-1}
\usepackage{amsmath}
\usepackage{amssymb}
\usepackage{graphics}
\usepackage{epsfig}

\begin{document}

\title{Critical current density and vortex pinning mechanism of K$_{x}$Fe$%
_{2-y}$Se$_{2}$ with S doping}
\author{Hechang Lei and C. Petrovic}
\affiliation{Condensed Matter Physics and Materials Science Department, Brookhaven
National Laboratory, Upton, NY 11973, USA}
\date{\today}

\begin{abstract}
We report critical current density $J_{c}$ in K$_{x}$Fe$_{2-y}$Se$_{2-z}$S$%
_{z}$ crystals. The $J_{c}$ can be enhanced significantly with optimal S
doping (z = 0.99). For K$_{0.70(7)}$Fe$_{1.55(7)}$Se$_{1.01(2)}$S$_{0.99(2)}$
the weak fishtail effect is found for H$\Vert $c. The normalized vortex
pinning forces follow the scaling law with maximum position at 0.41 of
reduced magnetic field. These results demonstrate that the small size normal
point defects dominate the vortex pinning mechanism.
\end{abstract}

\maketitle

\section{Introduction}

Since the discovery of LaFeAsO$_{1-x}$F$_{x}$ (FeAs-1111 type) with $T_{c}$
= 26 K,\cite{Kamihara} intensive studies have been carried out in order to
understand the superconducting mechanism, explore new materials and possible
technical applications. Among discovered iron-based superconductors,
FeAs-1111 materials and AFe$_{2}$As$_{2}$ (A = alkaline or alkaline-earth
metals, FeAs-122 type) exhibit high upper critical fields ($\mu _{0}H_{c2}$)
and good current carrying ability which are important for energy
applications.\cite{Hunte}$^{-}$\cite{Yang H} On the other hand, even though
FeCh (Ch = S, Se, and Te, FeCh-11 type) materials have nearly isotropic high
$\mu _{0}H_{c2}$ and considerable critical current density,\cite{Lei HC1}$%
^{,}$\cite{Yadav} their relatively low $T_{c}$ when compared to FeAs-1111
and FeAs-122 superconductors is a serious disadvantage. Recently, A$_{x}$Fe$%
_{2-y}$Se$_{2}$ (A = K, Rb, Cs, and Tl, AFeCh-122 type) materials attracted
much attention due to rather high $T_{c,onset}$ ($\sim $ 32 K), and $\mu
_{0}H_{c2}$($\sim $ 56 T for H$\parallel $c at 1.6 K).\cite{Guo}$^{,}$\cite%
{Mun} However, preliminary studies indicate that the critical current density in K$%
_{x}$Fe$_{2-y}$Se$_{2}$ is lower than in other iron based superconductors.\cite%
{Lei HC2}$^{,}$\cite{Hu RW} Therefore, it is important to explore pathways
for the critical current density $J_{c}$ enhancement in AFeCh-122 compounds.

In present work, we report the enhancement of critical current density\ and
vortex pinning mechanism in K$_{x}$Fe$_{2-y}$Se$_{2-z}$S$_{z}$ single
crystals. Point defect pinning dominates the vortex pinning mechanism
whereas critical current density is maximized for z = 0.99(2).

\section{Experiment}

Details of crystal growth and structure characterization were reported in
previous work.\cite{Lei HC2}$^{,}$\cite{Lei HC3} Crystals were polished into
rectangular bars and magnetization measurements were performed in a Quantum
Design Magnetic Property Measurement System (MPMS-XL5) up to 5 T. The
average stoichiometry and homogeneity of samples were determined by
examination of multiple points using an energy-dispersive x-ray spectroscopy
(EDX) in a JEOL JSM-6500 scanning electron microscope.

\section{Results}

\begin{figure}[tbp]
\centerline{\includegraphics[scale=0.4]{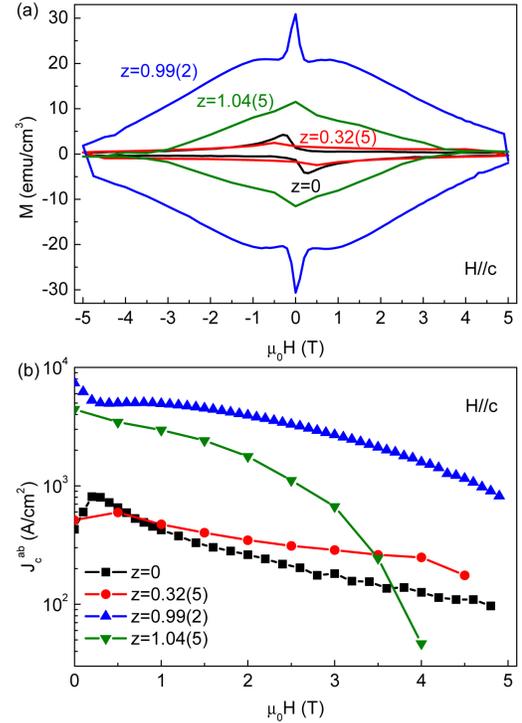}} \vspace*{-0.3cm}
\caption{(a) Magnetization hysteresis loops of K$_{x}$Fe$_{2-y}$Se$_{2-z}$S$%
_{z}$ at 1.8 K for H$\Vert $c. (b) Superconducting critical current
densities $J_{c}^{ab}(\protect\mu _{0}H)$ determined from magnetization
measurements using the Bean model.}
\end{figure}

Fig. 1(a) shows magnetization hysteresis loops (MHLs) of K$_{x}$Fe$_{2-y}$Se$%
_{2-z}$S$_{z}$ at 1.8 K for H$\Vert $c with field up to 5 T. The shapes of
MHLs for all of samples are typical of type-II superconductors. However, for
different S doping, they exhibit different flux pinning behavior. For low S
doping (z = 0 and z = 0.32), the MHLs are asymmetric. This asymmetry
suggests that the bulk pinning is small and that the influence of the
surface barrier is important.\cite{Pissas}$^{,}$\cite{Zhang L} On the other
hand, for higher S doping (z = 0.99 and z = 1.04), the shapes of MHLs\ are
symmetric indicating that the bulk pinning is dominant. For K$_{0.70(7)}$Fe$%
_{1.55(7)}$Se$_{1.01(2)}$S$_{0.99(2)}$ crystal, a small fishtail hump
appears at 0.8 T, similarly to FeAs-122 single crystals.\cite{Yang H}$^{,}$%
\cite{Sun DL}$^{-}$\cite{Prozorov} We determine the critical current density
from the Bean model.\cite{Bean}$^{,}$\cite{Gyorgy} For a
rectangularly-shaped crystal with dimension c $<$ a $<$ b, when H$\Vert $c,
the in-plane critical current density $J_{c}^{ab}(\mu _{0}H)$ is given by

\begin{equation}
J_{c}^{ab}(\mu _{0}H)=\frac{20\Delta M(\mu _{0}H)}{a(1-a/3b)}
\end{equation}

where a and b (a $<$ b) are the in-plane sample size in cm, $\Delta M(\mu
_{0}H)$ is the difference between the magnetization values for increasing
and decreasing field at a particular applied field value (measured in emu/cm$%
^{3}$), and $J_{c}^{ab}(\mu _{0}H)$ is the critical current density in A/cm$%
^{2}$. From Fig. 1(b), it can be seen that the $J_{c}^{ab}(\mu _{0}H)$ shows
small increase at high field region for z = 0.32 when compared to z = 0
sample. On the other hand, it is enhanced about one order of magnitude for z
= 0.99 in the whole magnetic field range. For higher S content, the $%
J_{c}^{ab}(0)$ is still much larger than in pure K$_{0.64(4)}$Fe$_{1.44(4)}$%
Se$_{2.00(0)}$, but the $J_{c}^{ab}(\mu _{0}H)$ at high fields is smaller.
It should be noted that the $T_{c}$ decreases significantly when z $>$ 0.32.
When compared to z = 0 ($T_{c,onset}$ = 33.0 K), z = 0.99 crystal has $%
T_{c,onset}$ = 24.6\ K, whereas z= 1.04 has $T_{c,onset}$ = 18.2 K.\cite{Lei
HC3} Therefore, sample with z = 0.99 exhibits the best performance and we
studied its field dependence of magnetization and critical current density
in detail. K, Fe, Se and S are uniformly distributed in z = 0.99 crystal
(Fig. 2), as is the case with all crystals we investigated.

\begin{figure}[tbp]
\centerline{\includegraphics[scale=0.4]{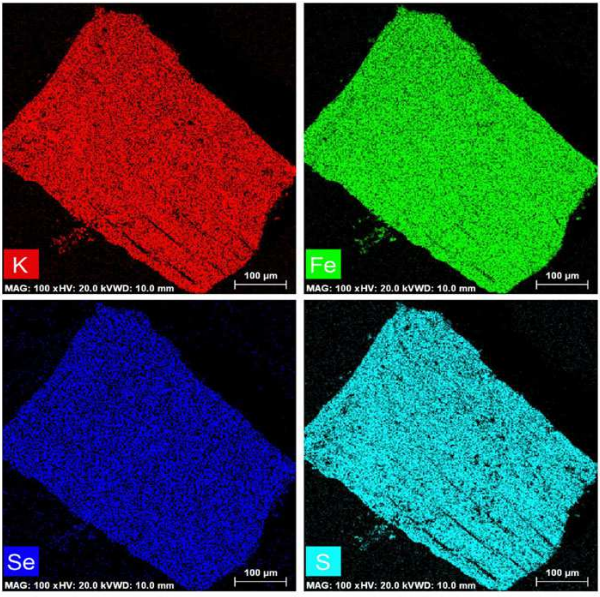}} \vspace*{-0.3cm}
\caption{EDX mapping of K$_{0.70(7)}$Fe$_{1.55(7)}$Se$_{1.01(2)}$S$%
_{0.99(2)} $. (Scale bar is 0.1 mm.)}
\end{figure}

\begin{figure}[tbp]
\centerline{\includegraphics[scale=0.72]{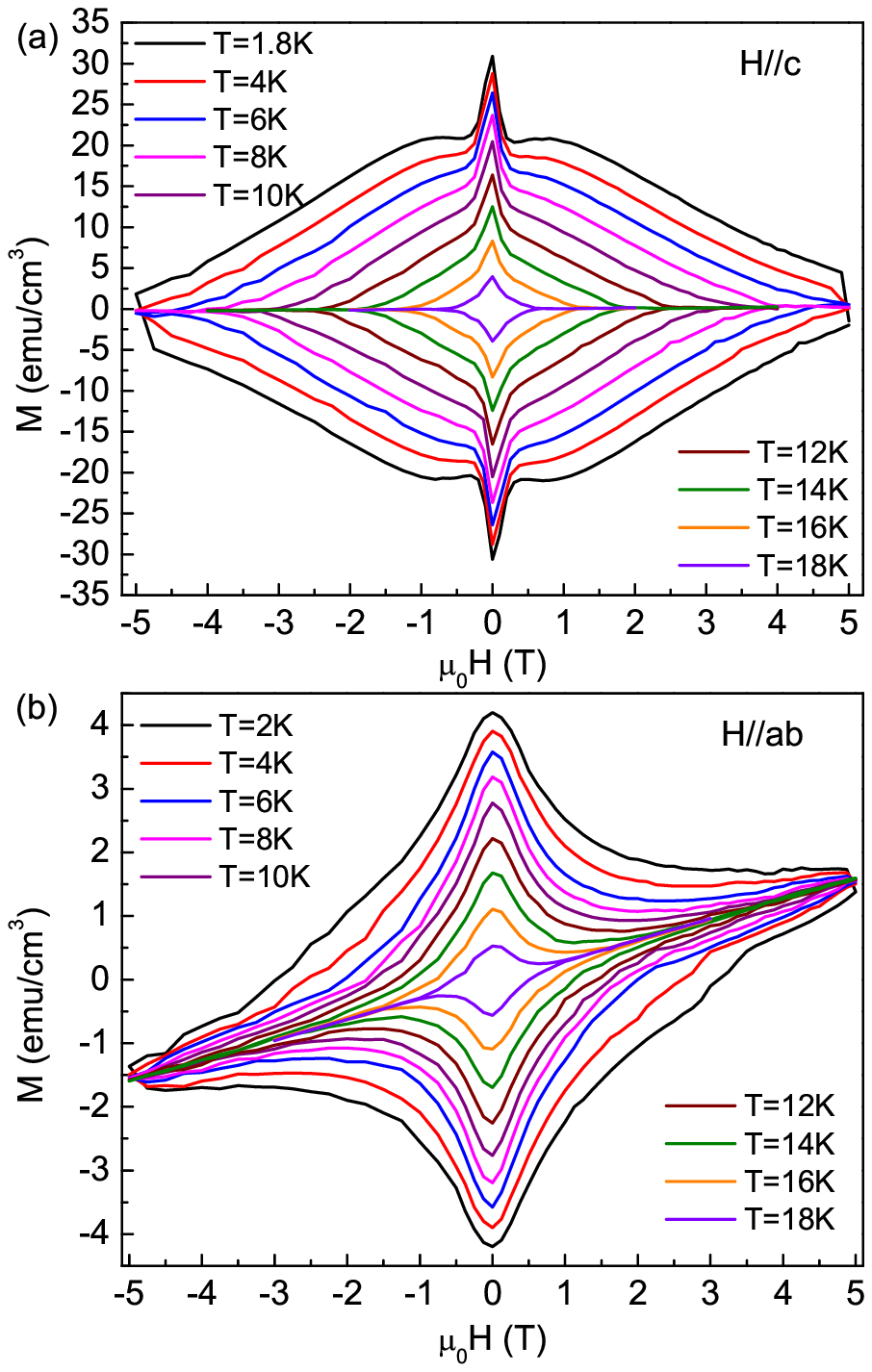}} \vspace*{-0.3cm}
\caption{MHLs of K$_{0.70(7)}$Fe$_{1.55(7)}$Se$_{1.01(2)}$S$_{0.99(2)}$ for
(a) H$\Vert $c and (b) H$\Vert $ab.}
\end{figure}

Fig. 3 shows the MHLs of crystal with z = 0.99 for both field directions.
The fishtail effect\ is only observed for H$\Vert $c. It diminishes
gradually with increasing temperature. Similar behavior has also been seen
in BaFe$_{2-x}$Co$_{x}$As$_{2}$,\cite{Prozorov} suggesting anisotropic flux
pinning.\cite{Sun DL} On the other hand, linear $M(\mu _{0}H)$ background
exists for both field directions, being more obvious for H$\Vert $ab. This
is also observed in pure K$_{0.64(4)}$Fe$_{1.44(4)}$Se$_{2.00(0)}$.\cite{Lei
HC2} The slope of this background for crystal with z = 0.99 is nearly the
same as in pure material, suggesting that high temperature magnetism changes
little with z for z $\leqslant $ 0.99. The linear $M(\mu _{0}H)$ background
has no effect on the calculation of $\Delta M(\mu _{0}H)$, and is due to
incomplete superconducting volume fraction of the crystals used in this
study. We will discuss the effects of electromagnetic granularity in the
next section.

\begin{figure}[tbp]
\centerline{\includegraphics[scale=0.72]{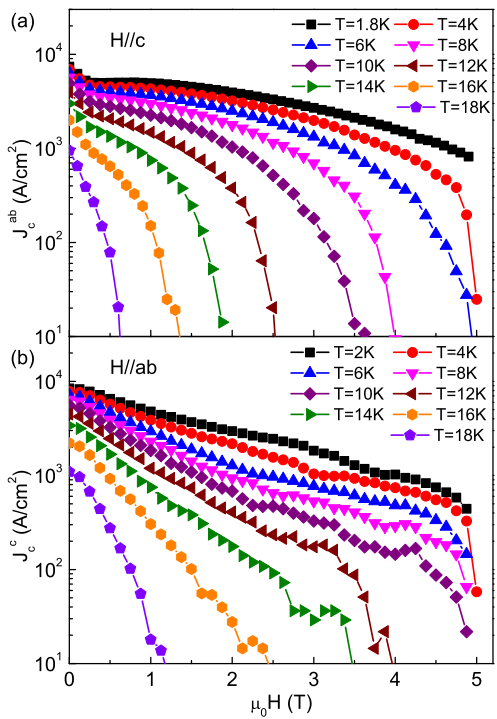}} \vspace*{-0.3cm}
\caption{Magnetic field dependence of superconducting critical current
densities (a) $J_{c}^{ab}(\protect\mu _{0}H)$ and (b) $J_{c}^{c}(\protect\mu %
_{0}H)$ for K$_{0.70(7)}$Fe$_{1.55(7)}$Se$_{1.01(2)}$S$_{0.99(2)}$.}
\end{figure}

The $J_{c}^{ab}(\mu _{0}H)$ is calculated using eq. (1) for H$\Vert $c and
shown in Fig. 4(a)The evaluation of critical current density becomes more
complex for H$\Vert $ab, since there are two different contributions. One is
vortex motion across the planes, $J_{c}^{c}(\mu _{0}H)$, and the other is
vortex motion parallel to the planes, $J_{c}^{\Vert }(\mu _{0}H)$. Usually, $%
J_{c}^{\Vert }(\mu _{0}H)\neq J_{c}^{ab}(\mu _{0}H)$. Assuming $a,b\gg
c/3\cdot J_{c}^{\Vert }(\mu _{0}H)/J_{c}^{c}(\mu _{0}H)$,\cite{Gyorgy} we
obtain $J_{c}^{c}(\mu _{0}H)\approx 20\Delta M(\mu _{0}H)/c$. The calculated
$J_{c}^{c}(\mu _{0}H)$ is shown in Fig. 3(b).

The $J_{c}^{ab}(0)$ and $J_{c}^{c}(0)$ are 7.4 and 8.4$\times $10$^{3}$ A/cm$%
^{2}$ at 1.8 K and 2 K, respectively. Even though they are still smaller
than in other iron pnictide superconductors (where critical current
densities are usually above 10$^{5}$ A/cm$^{2}$ at 5 K),\cite{Sun DL}$^{,}$%
\cite{Tanatar} S doping significantly enhances the critical current density
when compared to pure K$_{0.64(4)}$Fe$_{1.44(4)}$Se$_{2.00(0)}$.\cite{Lei
HC2}$^{,}$\cite{Hu RW} It suggests that S doping introduces effective
pinning center and therefore enhances the $J_{c}$ for both field directions.
On the other hand, the ratio of $J_{c}^{c}(\mu _{0}H)$/$J_{c}^{ab}(\mu
_{0}H) $ is approximately 1 and is smaller than in BaFe$_{2-x}$Co$_{x}$As$%
_{2}$.\cite{Tanatar} The field dependence of $J_{c}^{c}(\mu _{0}H)$ is
somewhat weaker than $J_{c}^{ab}(\mu _{0}H)$. This could be related to the
layered structure of K$_{x}$Fe$_{2-y}$Se$_{2-z}$S$_{z}$.

\section{Discussion}

\begin{figure}[tbp]
\centerline{\includegraphics[scale=0.8]{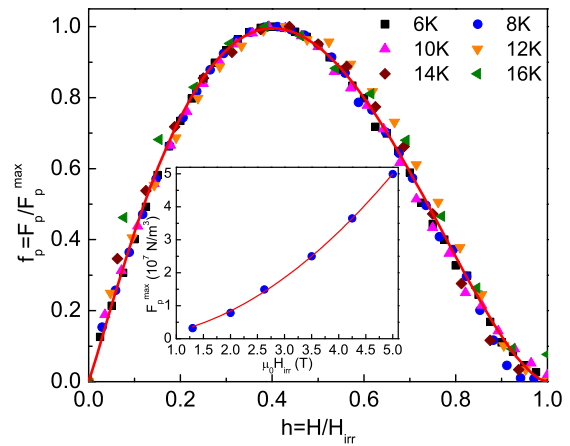}} \vspace*{-0.3cm}
\caption{Normalized flux pinning force $f_{p}=F_{p}/F_{p}^{\max }$ as a
function of reduced field $h=H/H_{irr}$ for K$_{0.70(7)}$Fe$_{1.55(7)}$Se$%
_{1.01(2)}$S$_{0.99(2)}$. Solid line represents the fitting curve using $%
f_{p}=Ah^{p}(1-h)^{q}$. Inset shows $F_{p}^{\max }$ as a function of $%
\protect\mu _{0}H_{irr}$. Solid line shows the fitting result obtained by
using $F_{p}^{\max }=A(\protect\mu _{0}H_{irr})^{\protect\alpha }$.}
\end{figure}

In order to gain more insight into the vortex pinning mechanism in K$%
_{0.70(7)}$Fe$_{1.55(7)}$Se$_{1.01(2)}$S$_{0.99(2)}$, we plot the normalized
vortex pinning force $f_{p}=F_{p}/F_{p}^{\max }$ as a function of the
reduced field $h=H/H_{irr}$ at various temperatures for H$\Vert $c (Fig. 5).
The pinning force $F_{p}$ was obtained from the critical current density
using $F_{p}=\mu _{0}HJ_{c}$, and $F_{p}^{\max }$ corresponds to the maximum
pinning force. The irreversibility field $\mu _{0}H_{irr}$ is the magnetic
field at which $J_{c}^{ab}(T,\mu _{0}H)$ is zero. It can be clearly seen
that the $f_{p}$ vs $h$ curves exhibit scaling behavior, independent of
temperature, suggesting dominance of single vortex pinning mechanism.
Scaling law $f_{p}\propto h^{p}(1-h)^{q}$ explains well our data.\cite%
{Dew-Hughes} The obtained parameters are $p$ = 1.10(1) and $q$ = 1.64(2).
The value of $h_{\max }^{fit}$ ($=p/(p+q)$) $\approx $ 0.40 is consistent
with the peak positions ($h_{max}^{\exp }\approx $ 0.41) of the experimental
$f_{p}$ vs $h$ curves at various temperatures. According to the Dew-Hughes
model for pinning mechanism,\cite{Dew-Hughes} the $h_{max}$ = 0.33 with $p$
= 1 and $q$ = 2 corresponds to small size normal point defects pinning. Our
results indicate that small normal point defects pinning dominates vortex
pinning mechanism. These defects could be related to distribution of S ions
on a submicron scale, similarly to FeAs-122 system.\cite{Yang H}$^{,}$\cite%
{Sun DL}$^{,}$\cite{Yamamoto} Moreover, the $F_{p}^{\max }$ can be fitted
using $F_{p}^{\max }=A(\mu _{0}H_{irr})^{\alpha }$ and we obtain $\alpha =$
1.94(3). This is consistent with the theoretical value ($\alpha $ = 2).\cite%
{Dew-Hughes}

Since our crystals are rather homogeneous (Fig. 2), the shape of M(H) for z
= 0 and z = 0.32 suggests some electromagnetic granularity similar to SmFeAsO%
$_{0.85}$F$_{0.15}$.\cite{Senatore} Since we used the full sample dimensions
in $J_{c}^{ab}$ and $J_{c}^{c}$ calculation, the values we obtained
represent the lower limit of bulk superconducting crystals. Indeed, very
recently Gao et al. reported the $J_{c}$ of K$_{x}$Fe$_{2-y}$Se$_{2}$ can be
enhanced significantly (about 1.7$\times $10$^{4}$ A/cm$^{2}$ at 5 K) using
one-step technique.\cite{Gao ZS} It implies that with S doping, the $J_{c}$
of AFeCh-122 might increase further if preparation process can be optimized.

\section{Conclusion}

In summary, we show an order of magnitude increases in $J_{c}$ by S doping
in K$_{x}$Fe$_{2-y}$Se$_{2}$. The optimum S content in K$_{x}$Fe$_{2-y}$Se$%
_{2-z}$S$_{z}$ single crystals is z = 0.99. For the optimally doped sample,
the weak fishtail effect is observed when H$\parallel $c. The analysis of
vortex pinning force indicates that the dominant pinning sources are small
size normal point defects which could originate from distribution of doped
S. The results demonstrate that by further optimizing the vortex pinning
force, higher values of $J_{c}$ could be achieved, raising the prospects for
technical applications of AFeCh-122 compounds.

\section{Acknowledgements}

We thank John Warren for help with SEM measurements. Work at Brookhaven is
supported by the U.S. DOE under Contract No. DE-AC02-98CH10886 and in part
by the Center for Emergent Superconductivity, an Energy Frontier Research
Center funded by the U.S. DOE, Office for Basic Energy Science.

\end{document}